\documentclass[conference]{IEEEtran}
\IEEEoverridecommandlockouts
\usepackage{cite}
\usepackage{amsmath,amssymb,amsfonts}
\usepackage{algorithmic}
\usepackage{graphicx}
\usepackage{textcomp}
\usepackage{xcolor}
\def\BibTeX{{\rm B\kern-.05em{\sc i\kern-.025em b}\kern-.08em
    T\kern-.1667em\lower.7ex\hbox{E}\kern-.125emX}}

\usepackage{flushend}

\usepackage{tabularx}
\newcolumntype{Y}{>{\centering\arraybackslash}X}

\graphicspath{{images/}}
\usepackage[normalem]{ulem}
\useunder{\uline}{\ul}{}

\begin{document}
\title{Ensemble Methods for Multi-Organ Segmentation in CT series\\
\thanks{This work is supported by Italian Ministry of Health (grant number: GR-2019-12370739, project: AuToMI).}
}

\author{
\IEEEauthorblockN{Leonardo Crespi\IEEEauthorrefmark{1}\IEEEauthorrefmark{2}, Paolo Roncaglioni\IEEEauthorrefmark{1}, Damiano Dei\IEEEauthorrefmark{3}\IEEEauthorrefmark{4}, Ciro Franzese\IEEEauthorrefmark{3}\IEEEauthorrefmark{4}, Nicola Lambri\IEEEauthorrefmark{3}\IEEEauthorrefmark{4},\\ Daniele Loiacono\IEEEauthorrefmark{1},
Pietro Mancosu\IEEEauthorrefmark{4}, and Marta Scorsetti\IEEEauthorrefmark{3}\IEEEauthorrefmark{4}}
\smallskip
\IEEEauthorblockA{\IEEEauthorrefmark{1}Dipartimento di Elettronica, Informazione e Bioingegneria, Politecnico di Milano, Milan, Italy}
\IEEEauthorblockA{\IEEEauthorrefmark{2}Centre for Health Data Science, Human Technopole, Milan, Italy}
\IEEEauthorblockA{\IEEEauthorrefmark{3}Department of Biomedical Sciences, Humanitas University, Pieve Emanuele, Milan, Italy}
\IEEEauthorblockA{\IEEEauthorrefmark{4}Radiotherapy and Radiosurgery Department, IRCCS Humanitas Research Hospital, Rozzano, Milan, Italy}
}

\maketitle

\begin{abstract}
In the medical images field, semantic segmentation is one of the most important, yet difficult and time-consuming tasks to be performed by physicians. Thanks to the recent advancement in the Deep Learning models regarding Computer Vision, the promise to automate this kind of task is getting more and more realistic. However, many problems are still to be solved, like the scarce availability of data and the difficulty to extend the efficiency of highly specialised models to general scenarios. Organs at risk segmentation for radiotherapy treatment planning falls in this category, as the limited data available negatively affects the possibility to develop general-purpose models; in this work, we focus on the possibility to solve this problem by presenting three types of ensembles of single-organ models able to produce multi-organ masks exploiting the different specialisations of their components. The results obtained are promising and prove that this is a possible solution to finding efficient multi-organ segmentation methods. 
\end{abstract}

%
\section{Introduction and Related Works}
Radiation therapy (RT) is a prevalent treatment for various types of cancer. It involves administering high doses of radiation to eliminate cancer cells and shrink tumors. When devising an RT treatment plan, it is crucial to deliver an appropriate dose distribution to tumoral targets while minimizing exposure to nearby organs, also known as organs-at-risk (OARs). Hence, accurate delineation and segmentation of OARs are essential for effective treatment planning, a process that is both time-consuming and error-prone for human experts. Recent advancements in the field of Deep Learning suggest that it is feasible to automate this segmentation process, commonly referred to as \emph{Semantic Segmentation} in the Machine Learning literature~\cite{Long2015CVPR}.

Significant progress has been made in semantic segmentation of medical imaging over the past few years. While atlas-based solutions remain widely used in commercial environments, state-of-the-art approaches now typically employ deep learning techniques such as auto-encoders (AE), convolutional neural networks (CNN), fully convolutional networks (FCN), generative adversarial networks (GAN), and regional-CNN~\cite{lei2020deep,asgari2021deep}. 
A major challenge when applying Deep Learning to medical imaging segmentation is the limited availability of annotated data. Specifically, when focusing on OARs segmentation for whole-body treatments like total marrow irradiation~\cite{wong2020tmi}, acquiring a substantial amount of training data with annotations for all organs to segment is difficult~\cite{lei2020deep,asgari2021deep}.
A promising approach to address this issue involves using several smaller datasets, which include annotations for only one or a few of the OARs to segment, to train multiple single-organ segmentation models that can be merged later for multi-organ segmentation. Additionally, this approach allows for different data preprocessing (e.g., applying a unique look-up table) for each model and, therefore, for each OAR. In this paper, we explore this research direction, inspired by multi-modal segmentation architectures \cite{review_multimodal_fusion}, in which the segmentation of different medical imaging acquisition modes (e.g., CT scans and MRI) are combined.
Ensemble learning, in general, aims to combine several models to improve individual performances. Accordingly, previous works in the literature~\cite{10.1007/978-3-319-75238-9_38,LI2018650} investigated the application of ensemble methods to achieve better performance than single models. However, only a few works have focused on employing ensembles to address the lack of annotated data in multi-organ segmentation. The most notable example is the work of Fang et al.~\cite{9112221}, who proposed an ensemble approach to train a model on multiple datasets with partial annotations. 

In this study, instead, we concentrate on exploring three ensembling strategies to combine previously trained single-organ segmentation models for multi-organ segmentation tasks on CT series. Our experimental analysis includes comparing different data sources and architectures (Unet, SE-ResUnet, and DeepLabV3). Our results are promising and indicate that ensemble methods can generally outperform or, at the very least, achieve similar performance to both multi-organ segmentation models and the single models used to construct them, while also reducing development effort.

\section{Proposed Solutions} \label{sec:methodology}%
	Three different methods have been developed and tested. The leitmotiv of this work is to use multiple binary segmentation models trained on single organs in ensembles, in order to exploit the previously acquired high specialisation to tackle a broader problem. Inspiration for the methods described in this section has often been drawn from multimodal multi-class segmentation scenarios \cite{review_multimodal_fusion}, with the difference that single-organ models have been used.
	
	\subsection{Binary models} \label{sub:binary}%
	
		As mentioned, the basic components of the methods proposed are binary models, each trained and specialised in the segmentation of one of six organs (right and left lung, heart, trachea, esophagus, spinal cord), the ones available in the StructSeg dataset (further described in \ref{sub:datasets}). Three different binary models for each organ have been trained, using some of the most popular architectures from the literature:
		\begin {itemize}%
			
			\item{\textbf{U-Net}}: introduced in \cite{unet}, is one of the most popular semantic segmentation architectures; it has a symmetric structure featuring an encoder, which downsamples the input extracting condensed features which represent the original data, and a decoder, responsible to upsample these features to create the desired output-the binary segmentation mask, in this case; skipped connections link the corresponding levels of downsampling-upsampling, respectively, of the encoder and the decoder so to reduce information loss and retain the original image structure. 
			\item{\textbf{SE-ResUNet}}: described in \cite{cascaded_seresunet} and inspired by ResUNet \cite{diakogiannis2020resunet} (which is a modification of U-Net using residual blocks \cite{he2016deep}), employs squeeze-and-excitation (SE)\cite{hu2018squeeze} blocks to enable the network to perform internal channel-wise features recalibration. It has been proven to be valuable in similar scenarios. 
			\item{\textbf{DeepLabV3}}: is another state-of-the-art architecture for semantic segmentation presented in \cite{deeplabv3}; once again, the network is composed by an encoder and a decoder, even though this architecture doesn't stem from the original U-Net; it relies, in the encoding part, on Spatial Pyramid Pooling \cite{he2015spatial}, which allows pooling while keeping the same spatial level feature together, and Atrous Convolutions \cite{he2015spatial}, able to gather information from distant pixels in the image, while also lightening the training process. 
		\end{itemize}
		
	\subsection{Baseline: Argmax ensemble} \label{sub:argmax}%
		The first and simplest ensemble method relies on a heterogeneous pool of models only and consists in assigning to each pixel of the resulting image the prediction from the most confident positive model above a certain threshold. The operation is an \emph{argmax} on C classes for each $(x, y)$ pixel on the set of the masks M:
		
		\begin{equation} \label{eq:argmax}%
			\underset{C}{argmax}M_c(x, y)
		\end{equation}
		
		The outputs of single models are stacked, then a sigmoid function is applied, it is thresholded, and the argmax is computed; the method requires no other computational effort than the forward pass in the networks and the result is a multi-class segmentation mask. The natural consequence is that the ensemble is strictly as good as the single models involved, as no further training is required. This ensemble method, together with multiclass single models, has been used as a baseline for the experiments.

    \begin{figure}
        \centering
        \includegraphics{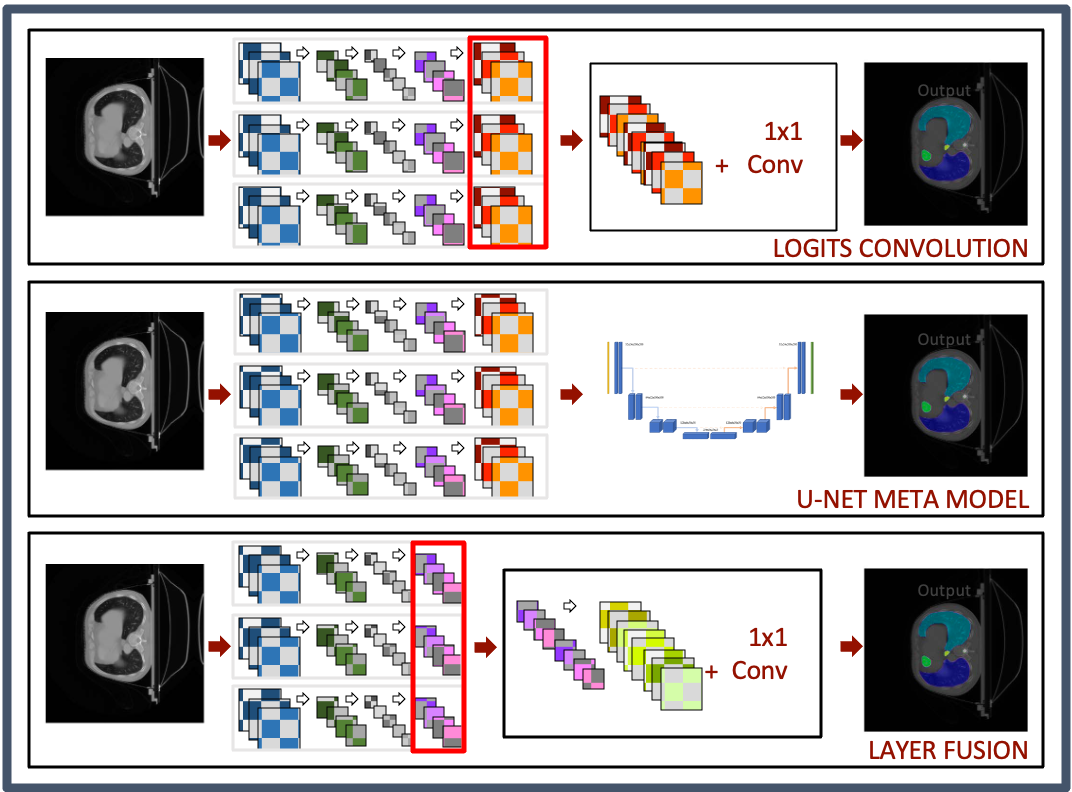}
        \caption{Schemes representing the three methods used to ensemble binary models. The coloured blocks represent the downsampling and upsampling steps, common in the networks specialised in segmentation, and do not have any intention of precisely representing the models used but rather of showing intuitively how the ensembles are assembled. }
        \label{fig:scheme}
    \end{figure}
		
	\subsection{Logits Convolution} \label{sub:e1}%
		The second method exploits the logits of the single models. It still relies heavily on their performance, but in this case, supplementary training is necessary: a 1x1 Convolutional layer combines the concatenated outputs of the binary models, reducing the feature maps to the number of labels while learning how to weight the single model, which might give it an edge on the simpler argmax; in the worst case scenario, the expected result is that the added layer learns to take the maximum value. Also, it is possible to analyse the weight assigned to a certain model for a single class, understanding more about the decision process. After the convolution, the resulting feature maps are thresholded and the final output is a multi-label mask. In the first row of \ref{fig:scheme}, this method is schematised. 
		
	\subsection{Meta U-Net} \label{sub:e2}%
        Multiple authors proposed with success cascaded models for segmentation of ROIs in the medical imaging field; in \cite{guo2021effective}, a method using two CNN put in sequence is shown, with the first extracting a piece of rougher information, like the bounding box containing the ROI, while the second actually segments the target from within the bounding box; it obtained very promising results on CT images, chest x-rays, and retinal images. Similarly, but for a different application, in \cite{christ2016automatic} the authors developed two cascaded fully convolutional networks to segment the liver first, and subsequently, lesions in it looking in the liver area only, all from CT scans. In this other work by Roth et al.,  \cite{roth2018application}, the authors propose a two-step process using CNNs to pass from a coarse segmentation to a finer one, using the input of the upper-level net to expand the information given to the second. In general, methods using a two steps system proved to be rather successful in the field; this third method stems from this consideration and uses the binary models up to the logits once again: the logits maps are stacked and fed to a second model, the meta-model, which is then trained to learn the actual segmentation map from them; being a segmentation problem, the meta-model of choice, in this case, is a U-Net \cite{unet}, which is able to take the concatenated feature maps as input. The second row in figure \ref{fig:scheme} shows a scheme of the architecture. An important remark is that during the training of the meta-model, the weights of the previous one are not updated. 
		
	\subsection{Layer fusion} \label{sub:e3}%
		In the last method, binary models are merged in their last layers before the logits, obtaining a more complex single model with multiple input branches. For what concerns U-Net and SE-ResUnet nets, the last layer is the one following the last transposed convolution/upsampling, right before the final 1x1 convolution; in DeepLabV3, the last upsampling layer has been kept, in order to have coherent features that can be concatenated. Hence, as shown in the third row of figure \ref{fig:scheme}, features from different binary models are concatenated, so that one would get 64 feature maps from U-Net and SE-ResUnet and 256 from DeepLabV3. Contrary to the previous methods, the weights for this last concatenated layers are not frozen during the training of the remainder of the model. A 1x1 convolutional layer follows the concatenation, similarly as before, to produce a number of logit maps corresponding to the number of classes; thresholding is then applied to obtain the final multilabel mask. 

\section{Experiments and Results}%
     In order to assess the capabilities of the ensembles proposed in this work, we carried out multiple experiments with the objective to simulate multiple scenarios that are likely to arise in the real world; this is aimed to provide a comprehensive overview of the strength and weaknesses of such an approach and also to highlight where this system may be useful and where this may not.
	\subsection{Datasets} \label{sub:datasets}%
		The data used to build and test models come from two popular public datasets specifically built for segmentation tasks in the context of two different challenges: \emph{Structseg}, from the \emph{Automatic Structure Segmentation for Radiotherapy Planning Challenge}, part of MICCAI 2019, and SegTHOR, from the challenge 4 of the IEEE International Symposium on Biomedical Imaging \cite{lambert2020segthor}.
		
		\begin{itemize}
			\item{\textbf{StructSeg}}: a collection of 50 thoracic CT scans from 50 different patients, annotated by expert oncologists with six labels representing six OARs during radiotherapy treatments for lung cancer (right and left lung, heart, trachea, esophagus, spinal cord).
            The CT scans come in a size of 512 x 512 pixels and their in-plane resolution is 1.2 mm x 1.2 mm. The number of slices in the scans varies between 80 to 127, and the z-resolution is 5 mm.
			\item{\textbf{SegTHOR}}: thoracic CT scans of patients affected by lung cancer or Hodgkin's Lymphoma, used to prepare radiotherapy treatments. The dataset contains 40 scans with 4 OARs annotated by an expert: heart, esophagus, trachea, and aorta. The CT scans come in a size of 512 x 512 pixels and their in-plane resolution varies between 0.90 mm and 1.37 mm per pixel, depending on the patient. The number of slices in the scans varies between 150 to 284, and the z-resolution lies between 2 mm and 3.7 mm. 
		\end{itemize}
		
	\subsection{Preprocessing and Data Augmentation} \label{sub:preprocessing}%
		In DL, it is common to use some light preprocessing steps in order to prepare the data for training, which will then be repeated for every data used at inference time and whenever the model is used again in the future; the aim is to make the data coherent while also improving the training process and possibly the results. Data augmentation is usually necessary on small datasets, as it has been demonstrated that training on a larger sample size yields better final results \cite{sun2017revisiting}; also, it makes models more robust to variations; similarly to preprocessing, there are some common techniques adopted to perform this, carefully trying to avoid as much as possible the introduction of bias. The following techniques have been used:

        \begin{table}
            \caption{Average scores over the labels for experiment 1.}
            \centering
            \begin{tabular}{||c||c|c|c|c|c||}
                \hline
                                    	& Lungs	    	        & Heart		                 & Esophagus 	    & Trachea      & Sp. Chord\\
                \hline \hline
                Width       			& 1500                 & 350                         & 300              & 1200         & 600   \\
                \hline
                Level       	       	& -600                 & 50                          & 80               & -440         & 0 \\
                \hline
            \end{tabular}
            \label{table:HU}
        \end{table}

		\begin{itemize}
			\item{\textbf{Lookup table (LUT) }}: is a technique used to preprocess CT  images to enhance their visualization and facilitate diagnosis.
            CT imaging involves acquiring a series of X-ray images of the body from different angles, which are then reconstructed into a 3D image using computer algorithms. The resulting CT images are often displayed in grayscale, where each pixel's intensity represents the attenuation of X-rays passing through the corresponding part of the body. The range of these intensities, expressed in Hounsfield Units, a dimensionless unit universally used in CT imaging, is [-1024, 3071] \cite{PREIM201415}, hence 12 bits of information. Windowing the range of values represented in an image on a smaller range, compressed to 8 bits, makes it easier to highlight a specific organ structure \cite{curry1990christensen}. The windows used, expressed in width (WW) and level (WL), are summarised in table \ref{table:HU}.
			\item{\textbf{Normalisation}}: the case for normalisation is given by the fact that values with a specific physical meaning, in particular regarding medical imaging, can be affected by the equipment or the different conditions of acquisition, therefore posing the risk of affecting the results, especially in a case where data come from different sources with the respect to the training set. To avoid this, pixels are normalised to the fixed range [0, 1].
			\item{\textbf{Center cropping}}: it has been used to speed up and lighten the training process because the actual informative and relevant area of the slices is limited to the central portion; therefore it is useless to analyse the whole 512x512; the optimal value found to keep all the relevant parts has been set to 320x320 on both datasets. 
            \item{\textbf{Geometric transformations}}: grid distortion, random rotations, and elastic transformation have been used for data augmentation; these methods allow for nondestructive manipulations of the original data and increase the sample size for training. They are not included in the test set; the transformations are applied to both the image and the mask so that the pixels still correspond. 
		\end{itemize}
		
	\subsection{Training} \label{sub:training}%
		 Training has been performed in two steps: single models first, of course, and ensembles later. One binary model with each of the architectures considered in \ref{sub:binary} has been trained on each OAR in a supervised fashion where the only considered mask is the one corresponding to that OAR. To train the ensembles, the parameters of the binary models have been kept frozen and only the ensemble parameters have been updated, as specified in section \ref{sec:methodology}. A composite loss function, comprising a Dice Score (DS) Loss term \cite{sudre2017generalised} and a Cross-entropy (CE) loss term, equally weighted, has been used, as suggested in \cite{ma2021loss} to take advantage of the characteristics of both; in literature \cite{cascaded_seresunet}, non-uniform weights for the different labels for the CE terms have been tested but after experiments with it, it has been decided to keep them uniform as they did not yield any advantage; an adaptive optimiser is the choice for the gradient descent algorithm (Adam, \cite{kingma2014adam}). Early stopping technique \cite{zhang2021understanding} helped improve the efficiency of the process, together with a progressive decrease of the learning rate (starting at $10^{-3}$ and exponentially decreased as in \cite{smith2017cyclical}) with reductions on plateau phases encountered in the validation loss curves, computed through a moving average to smooth the plot and take into consideration the past steps. 
		 
		 Pytorch 1.10 \cite{NEURIPS2019_9015} is the framework used to build and train the model in Python while the computational duties have been carried out on a machine equipped with an NVIDIA Quadro P4000 GPU with 8GB of memory for the binary models, and with an NVIDIA GeForce GTX 1060 with 6GB of memory for the ensembles.
	
	\subsection{Evaluation metrics} \label{sub:metrics}%
		Two main metrics have been chosen to be shown in this work to evaluate the performances of the models: the \emph{Dice Similarity Coefficient} (DSC) and the \emph{Hausdorff Distance 95} (HD95). The former \cite{taha2014formal} allows us to evaluate how much the mask overlaps to the GT; it assumes values in the range [0, 1] where 0 means no overlap and 1 identifies a perfect match. The HD95, differently, measures the distance between the two sets of points, highlighting how close the predicted mask is to the GT; the HD95 indicates that the 95 percentile is considered for this metric, avoiding possible outliers. These two metrics are fairly common in segmentation works and provide with a meaningful representation of the performance. 
		
	\subsection{Experiments} \label{sub:experiments}%
		Five different experiments have been run in order to assess the capabilities of the different ensemble methods described in the paper. The ensembles have been assembled using only the best binary network for each organ, according to their validation loss during training; the choices have been: \emph{SE-ResUNet} for left and right lung, esophagus, and spinal cord, and \emph{DeepLabv3} for heart and trachea. \emph{U-Net} has not performed best in any of the tasks considered for binary models. 
		
		\subsubsection{Baseline} 
			
			\begin{table}
            \caption{Average scores over the labels for the baseline models. \emph{U-Net, Se-ResUnet, DeepLabV3} are all multiclass models used as baseline.}
			\centering
			\begin{tabular}{||l||c|c|c|c||}
				\hline
				\textbf{Model}        	& \textbf{Dice}		& \textbf{Precision}		& \textbf{Recall} 	& \textbf{HD95 (mm)}		\\
				\hline \hline
				U-Net			& 0.824			              & \textbf{0.870}			& 0.805			             & 4.573			\\
				\hline              
    				SE-ResUNet		& 0.817			              & 0.840				     & 0.818			      & 6.347			\\
				\hline                          
    				DeepLabV3		& 0.851			               & 0.856				     & 0.854			      & 2.453			\\
				\hline      
				Argmax			& \textbf{0.878}			    & 0.858				     & \textbf{0.908}			  & \textbf{2.445}		\\
				\hline
			\end{tabular}
			\label{table:baseline}
			\end{table}		 
		 
			The Argmax ensemble has been used as an ensemble baseline because it represents the most confident output of the binary network and would be the same as always having the best binary network to segment each organ. In order to broaden the comparison pool, multiclass networks have been trained, one for each architecture considered; the process described in \ref{sub:training} still applies to this. Results obtained are shown in table \ref{table:baseline}; it is evident how the argmax ensemble has a clear edge on all the multiclass networks both in DSC and HD95.

		\subsubsection{Experiment 1 - Full Dataset} \label{subsub:ex1}

            \begin{table}
                \caption{Average scores over the labels for experiment 1.}
    			\centering
    			\begin{tabular}{||l||c|c|c|c||}
    				\hline
    				\textbf{Method}        	& \textbf{DICE}	    	& \textbf{PRECISION}		& \textbf{RECALL} 	& \textbf{HD95 (mm)}   \\
    				\hline \hline
    				Layer Fusion			& 0.874                 & 0.881                      & 0.875            & \textbf{2.260}\\
    				\hline
    				Logits Conc.	       	& \textbf{0.879}        & 0.869                       & \textbf{0.899}   & 2.294            \\
    				\hline
    				Meta U-Net      		& 0.866                 & \textbf{0.900 }            & 0.846             & 2.367             \\
    				\hline
    			\end{tabular}
    			\label{table:exp1}
            \end{table}		
                
			The first experiment consist in the straightforward training of the ensembles, followed by inference on the test dataset. The average results on the labels are shown in table \ref{table:exp1}. Looking at the DSC, it seems that no significant differences are noticeable with the respect to the baseline, with a performance that is actually even worse for the Meta U-Net by one percentage point, despite it showing the best precision. All the ensembles, in particular Layers Fusion and Logits Convolution, are able, however, to improve the HD95 score, meaning that the prediction is still more reliable, shape-wise, to the GT, than the baseline. %

        \begin{table}
            \caption{Average scores over the labels for experiment 2.}
			\centering
			\begin{tabular}{||l||c|c|c|c||}
				\hline
				\textbf{Method}        	& \textbf{DICE}	    	& \textbf{PRECISION}		& \textbf{RECALL} 	   & \textbf{HD95 (mm)}   \\
				\hline \hline
				Layer Fusion		    & \textbf{0.885}         & \textbf{0.912}            & 0.868               & 2.207 \\
                \hline  
				Logits Conc.	        & 0.884                  & 0.906                     &\textbf{0.872}       & \textbf{2.171} \\
                \hline
				Meta U-Net      	    & 0.877                 & 0.898                       & 0.870               & 2.292 \\
                \hline
			\end{tabular}
			\label{table:exp2}
        \end{table}		
			
		\subsubsection{Experiment 2 - Redundant branches} \label{subsub:ex2}
			Considering the single label scores (for space reasons, visible in the supplementary material), it is evident how some organs are more difficult to segment than others: in particular trachea and esophagus show the worst results both in terms of DSC and HD95. This is possibly due to the fact that these two organs are rather small compared to the others, therefore the positive/negative pixel ration is quite disadvantageous and the models struggle to correctly identify them. This second experiment tests the capabilities of the ensembles to improve on these targets by adding a second binary network specialised in the segmentation of such OARs: a supplementary DeepLabV3 for the esophagus and a supplementary U-Net for the trachea are then added. A clear improvement (table \ref{table:exp2}) is shown in both DSC and HD95 for all the ensembles, that now consistently score better than the baseline in the most relevant scores. Among the esembles, the meta model is still the worse, possibly due to a more difficult training process. %

        \begin{table}
            \caption{Average scores over the labels for experiment 3. The Argmax reported here is assembled with the binary networks trained on data from a single source.}
			\centering
			\begin{tabular}{||l||c|c|c|c||}
				\hline
				\textbf{Method}        & \textbf{DICE}	    	 & \textbf{PRECISION}		& \textbf{RECALL} 	& \textbf{HD95 (mm)}   \\
				\hline \hline
				Layer Fusion		    & \textbf{0.816}          & 0.838                     & \textbf{0.810}     & \textbf{4.412} \\  
                \hline          
				Logits Conc.	        & 0.760                    & 0.755                     & 0.771             & 21.761 \\
                \hline          
				Meta U-Net      	    & 0.782                    & \textbf{0.858}           & 0.761              & 7.366 \\
                \hline
                Argmax                  & 0.758                    & 0.748                     & 0.779             & 7.613 \\
                \hline
			\end{tabular}
			\label{table:exp3}
        \end{table}	
						
		\subsubsection{Experiment 3 - Differentiated sources} \label{subsub:ex3}
			In a real world scenario might happen that there are models available, but those are trained on different data with the respect to the ones available; of course transfer learning is an option, at least to try to lightly update the weights with some samples from the available data, but this is not always a viable possibility and it doesn't guarantee to produce reliable results; ensembles might help to solve this problem and with this experiment the aim is to simulate a scenario similar to the one described: binary models for heart, lungs and spinal cord have been trained on the StructSeg dataset only, while trachea and esophagus on SegThor, using the most succesful architectures of the previous experiments. The results, shown in table \ref{table:exp3}, obtained from the ensembles trained then on both the datasets together, are largely underwhelming, with consistently and significantly worse scores in all the categories with the respect to other scenarios; however, the Layer Fusion actually scores significantly better than the argamx ensemble obtained with this same models, proving to be a more viable approach. The general performance drop was to be expected because, first of all, of the negatively affected sample size, which cripples the performances of the binary networks, but the generalisation to new data is inherently worse for the same reason.

        \begin{table} 
            \caption{Average scores over the labels for experiment 4.}
			\centering
			\begin{tabular}{||l||c|c|c|c||}
				\hline
				\textbf{Method}        	& \textbf{DICE}	    	& \textbf{PRECISION}		& \textbf{RECALL} 	& \textbf{HD95 (mm)}   \\
				\hline \hline
				Layer Fusion		      & 0.880                   & \textbf{0.891}         & 0.876               & \textbf{2.075} \\  
                \hline
				Logits Conc.	           & \textbf{0.881}         & 0.869                  & \textbf{0.902}      & 2.159 \\
                \hline
				Meta U-Net      	       & 0.867                  & 0.889                  & 0.860               & 2.284 \\
                \hline                  
			\end{tabular}
			\label{table:exp4}
        \end{table}	
			
        \subsubsection{Experiment 4 - More redundancy} \label{subsub:ex4}
            In order to study the behaviour of the different ensembles methods with more diversity of models, in the fourth experiment, together with the binary networks, a multiclass one (\emph{DeepLabV3}) has been included, expecting a better overall performance because of the richer pool of segmentators included. 
            Following on the previous one, once again the aim is to test the ensembles changing the diversity of the models and thus, this time, including a multiclass network (a multiclass \emph{DeepLabV3}). Training and testing has been carried out in the same way as for experiments 1 and 2; this seems to yield the best results yet in terms of HD95 with the Layers Fusion while the DSC are on par with the ones from experiments 2, meaning that the addition of a multi-purpose network to an ensemble of specialised ones leads to only a slightly improved performance. However, in a real case scenario this approach might be more difficult as usually, not enough data is available to train both a multiclass network and the ensemble, therefore the increased burden necessary to add it might not be worth the performance increment. 

        \begin{table}
            \caption{Average scores over the labels for experiment 5.}
			\centering
			\begin{tabular}{||l||c|c|c|c||}
				\hline
				\textbf{Method}        & \textbf{DICE}	    	& \textbf{PRECISION}		& \textbf{RECALL} 	& \textbf{HD95 (mm)}\\
				\hline \hline
				Layer Fusion		  & 0.868                      & 0.866                     & 0.881	             & 2.366\\  
                \hline
				Logits Conc.	      & \textbf{0.879}             & 0.868                     & \textbf{0.899}	     & \textbf{2.298}\\
                \hline
				Meta U-Net      	  & 0.876                      & \textbf{0.903}            & 0.857 	             & 2.309\\
                \hline
			\end{tabular}
			\label{table:exp5}
        \end{table}	
			
		\subsubsection{Experiment 5 - Data scarcity} \label{subsub:ex5}
			The last experiments tries to emulate a scneario, likely to occur in the real world, were the data to train the ensemble is even scarcer than what's available for these challenges; using binary models as in experiment 1, because the hypothesis is that pre-trained models are available, the training set for the ensembles has been reduced to the 20\% of the initial one. The trend is the same showed in the previous tests, with a surprisingly low drop in performance, still on par with the baseline in the DSC but superior considering HD95. The Logits Convolution is the best performing ensemble, closely followed by the Meta U-Net, that in this scenario unexpectedly seems to be able to grasp the most out of the reduced sample size. 
			
		In general, from the results available, it seems to be possible to consider that the ensembles are generally superior than the baseline, considering both the argmax and the multiclass networks, and provide a powerful and promising tool to combine highly specialised single models in ensembles for multi-label segmentation, so that the whole becomes more than the sum of its parts. The solutions shown seems to be also robust to variations in the sample size and able to consistently score better as more specialised models are available. The most obvious drawback of such an approach is that binary models, specialised in the necessary task, have to be available, which is not always the case, but thanks to the increasing diffusion of public challenges, datasets and the efforts to share weights and models themselves through public repositories and so on, this is very likely to improve, to the point where it will be beneficial to combine them rather than to develop new ones.

\section{Conclusion}

In this study, we introduced ensembles of various Deep Neural Networks, each specialized in the segmentation of a single organ, with the goal of producing multi-organ segmentation masks. We tested three different methods inspired by multimodal segmentation ensembles and conducted several experiments to evaluate their strengths and weaknesses in different scenarios, attempting to mimic realistic real-world situations. The data used for training and testing the models and ensembles were sourced from two popular public challenges.

The results obtained are promising, as they suggest that the proposed solutions could be valuable systems for multi-organ segmentation tasks in situations where binary models specialized in single organs are available, and the data is not suitable for training high-performing networks from scratch. Nevertheless, a more in-depth research is required, involving testing different compositions of the ensembles, introducing additional data sources, and considering other anatomical regions and organs to determine if these methods can genuinely aid in real-world clinical practice. Furthermore, it is essential to test better-performing models, potentially already available within the community, to fully assess the realistic applicability of such an approach.

\bibliographystyle{IEEEtran} 
\bibliography{bibliography.bib}
\end{document}